\documentclass[12pt]{article}
\usepackage{amssymb}
\usepackage{amsmath}
\usepackage{graphicx}
\input{epsf.tex}
\begin{document}
\title{\vskip-2truecm{\hfill {\small CFNUL/00-03}} \\
\vskip 0.001truecm{\hfill {\small DF/IST-4.2000}} \\
\vskip-3mm
{\normalsize \hfill } \\
\vskip 5mm  
Gravitational instabilities in helicity-1 waves propagating
through matter in equilibrium}
\author{Lu\'{\i}s Bento\footnote{E-mail: lbento@fc.ul.pt } \\
{\normalsize \emph{Centro de F\'{\i}sica Nuclear da Universidade de Lisboa,}
} \\
{\normalsize \emph{Avenida Prof. Gama Pinto 2, 1649-003 Lisboa, Portugal,}} 
\\
{\ } \\
Jos\'{e} P. S. Lemos\footnote{E-mail: lemos@kelvin.ist.utl.pt} \\
{\normalsize \emph{CENTRA, Departamento de F{\'{\i}}sica, Instituto Superior
T\'{e}cnico,}} \\
{\normalsize \emph{Avenida Rovisco Pais 1,1049-001 Lisboa, Portugal, }} \\
{\normalsize \&} \\
{\normalsize \emph{Departamento de Astrof\'{\i}sica, Observat\' orio
Nacional-CNPq,}} \\
{\normalsize \emph{Rua General Jos\'e Cristino 77, 20921 Rio de Janeiro,
Brazil.}}}
\date{August, 2000}
\maketitle
\begin{abstract}
It is shown that the interaction of helicity-1 waves of gravity and
matter in a thin slab configuration produces new types of
instabilities. Indeed, a spin-2 helicity-1 mode interacts
strongly with the shear motion of matter.  This mode is unstable above
a critical wavelength, $\lambda _{c}=\sqrt{\pi c^{2}/2G\rho }$. This
should be compared with Jeans wavelength, $\lambda _{J}=\sqrt{\pi
{c_{s}^{2}}/G\rho }$, where $c_{s}$ is the sound speed. The two
instabilities are of course different. For the case analyzed, a plane
parallel configuration, Jeans instability appears through a density
wave perturbation, the material collapsing into a set of
plane-parallel slabs. On the other hand, the helicity-1 wave
instability induces a transverse motion in the fluid that tends to
shear in the material along the node of the perturbation.
 \newline
\strut \newline
\end{abstract}
\baselineskip=16pt
\newpage
Acoustic waves are affected by gravitational forces. One of the
consequences is Jeans instability \cite{jeans} (see also, e.g., \cite
{zeldovichnovikov,peebles}) produced above a certain critical
wavelength.  There is an electromagnetic counterpart in plasma density
waves. In the limit of large wavelengths their frequency square tends
to a finite positive value, and the plasma acquires the plasma
frequency, whereas in the Jeans case the frequency square turns
negative. This is noted by Weinberg \cite {weinberg72} but he does not
go further. Plasma density waves can be seen as longitudinal
electromagnetic waves which are not possible in vacuum but are present
in media with the plasma frequency having the role of a photon mass.
By the same token acoustic waves can be identified with longitudinal
waves of the gravity field but now, instead, they possess an imaginary
mass. Transverse electromagnetic waves, in addition to longitudinal
waves, are also modified in a plasma and acquire a mass equal to the
plasma frequency. The questions posed in this paper concern waves of
the gravity field, or wave metric perturbations, propagating in
matter. More specifically, we are interested in the following
question: are there waves which get a finite real mass or do
they get an imaginary mass leading to instabilities other than Jeans?

Waves of the gravity field can be classified according to their spin
and helicity.  Here we are interested in spin 2 waves which can have
helicities 2, 1 and 0.  Helicity-2 waves represent radiative
gravitational waves and are simply called gravitational waves.
Helicity-1 waves, as well as helicity-0, do not exist in vacuum but
can propagate in matter.  The propagation of waves of the gravity
field through matter has been studied in two main contexts, namely,
cosmological and astrophysical. Within the cosmological context these
waves appear as perturbations of the background universe
\cite{lifshitz} (see also \cite{grishchuk}) and there is now a growing
interest in the primordial gravitational wave background since it
seems possible to detect it in the future \cite{kolbturner,allen}.
Within the astrophysical context they have been studied mainly in
relation to stars (see e.g., \cite{kokkotas}). Usually one considers
these stars as either spherically or axisymmetric. The waves in the
star, i.e., perturbed quadrupolar or higher polar modes, appear as
convective velocity perturbations and density perturbations which
through the stellar boundary match with gravitational waves
propagating in vacuum \cite{detweiler,moncrief}. Still within the
astrophysical context, there has been another line of research, which
is the one that most concerns us here, analyzing non-dissipative \cite
{grishchuk,ehlers1,ehlers2} and dissipative propagation of waves
through matter \cite{polnarev,chesters,asseo76,galtsov,grishchuk}. The
results obtained so far confine themselves either to the short
wavelength limit or to transverse  helicity-2 gravitational wave
modes (e.g., \cite{ehlers1,asseo76}). In either case the interaction
of wave and matter is minute, and usually the wave propagates
undisturbed.

In this paper we want to go a step further, and reach the level where
the wavelength, $\lambda $, is of the order of or larger than the
background curvature, $\sim c/\sqrt{G\rho }$. This is not possible to
achieve for spherical or quasi-spherical geometries, since it implies
systems with a size of the order of their Schwarzschild
radius. However, extended finite thin plates either disk-like or
parallelepiped-like are amenable to the large $\lambda $ regime. In
order to progress we will study a specific case, namely a wave
propagating through a plane parallel slab. We are then able to find
some modes which propagate and truly interact with matter. The
transverse helicity-2 gravitational wave mode propagates freely
without interaction as has been observed in several papers (see
e.g. \cite {grishchuk,ehlers2}). On the other hand, a transverse
helicity-1 shear wave mode interacts strongly with the shear
motion of matter. Indeed, above a certain critical wavelength this mode
is unstable, as we will now show. 

Consider a fluid with proper energy density $\rho$, pressure $p$ and with a
given equation of state $p(\rho)$. Consider then it has suffered a small
departure from a state where it is at rest in a Minkowski space with
homogeneous density, $\rho_{0}$, and pressure, $p_{0}$. The fluid
energy-momentum tensor is ($c=1$) 
\begin{equation}
T_{\mu\nu}=(\rho+p)\,u_{\mu}u_{\nu}-p\,g_{\mu\nu} \;, \label{um}
\end{equation}
where $u_{\mu}$ denotes the 4-velocity vector. Greek indices $\mu,\, \nu$,
etc., take values $0$ (or $t$), and $i$, with $i=1,2,3$ (or $x,\, y, \, z$,
respectively). The metric $g_{\mu\nu}$ is denoted in the Minkowski limit as $
\eta_{\mu\nu}= \mathrm{diag} (1,-1,-1,-1)$. The fluid and the gravitational
field form a coupled system. The change of $T_{\mu\nu}$ due to a fluctuation
in the metric is determined by the conservation laws of energy and momentum, 
$\nabla^{\mu}T_{\mu\nu}=0$ and in turn, the gravitational field is
determined by the Einstein equations 
\begin{equation}
R_{\mu\nu}=8\pi G (T_{\mu\nu}-\tfrac{1}{2}g_{\mu\nu}T)\;. \label{dois}
\end{equation}
One has to distinguish between a static gravitational field, $\bar{g}
_{\mu\nu }=\eta_{\mu\nu}+\bar{h}_{\mu\nu}$, caused if not for anything else
by the fluid at rest, from the wave modes $h_{\mu\nu}$ defined
as 
\begin{equation}
g_{\mu\nu}=\eta_{\mu\nu}+\bar{h}_{\mu\nu}+h_{\mu\nu}\;. \label{tres}
\end{equation}
The field $\bar{h}_{\mu\nu}$ is related through Einstein equations, 
\begin{equation}
R_{\mu\nu}[\bar{h}]=8\pi G (\bar{T}_{\mu\nu}-\tfrac{1}{2} \bar{g}_{\mu\nu} 
\bar{T}) \;, \label{quatro}
\end{equation}
to the\ energy-momentum tensor, $\bar{T}_{\mu\nu}$, of the fluid in static
equilibrium under the external field $\bar{h}_{\mu\nu}$ itself. $\bar{T}
_{\mu\nu}$ is characterized by $\bar{\rho}$ and $\bar{p}$, which in turn
deviate slightly from $\rho_{0}$ and $p_{0}$, see below.

We adopt the de Donder gauge (see, e.g., \cite{mtw,thorne,schutz}), 
\begin{equation}
\partial^{\mu}h_{\mu\nu}-\frac{1}{2}\partial_{\nu}h_{\alpha}^{\alpha }=0\;,
\label{cinco}
\end{equation}
for both $h_{\mu\nu}$ and $\bar{h}_{\mu\nu}$. Here and in the following the
Minkowski metric $\eta_{\mu\nu}$ is used to raise and lower indices of $
h_{\mu\nu}$, $\bar{h}_{\mu\nu}$\ and $\partial_{\mu}$. In this gauge the
equations of energy-momentum conservation are in leading order in the
gravitational field 
\begin{equation}
(\eta^{\mu\alpha}-\bar{h}^{\mu\alpha}-h^{\mu\alpha})\,\partial_{\mu}
T_{\alpha\nu}=\frac{1}{2}T_{\mu\alpha}\,\partial_{\nu}(\bar{h}^{\mu\alpha
}+h^{\mu\alpha})\;. \label{seis}
\end{equation}
The limit $h_{\mu\nu}=0$ gives the conservation equations of $\bar{T}
_{\mu\nu }$. After subtracting that static contribution one obtains the
equations governing the variations $\tau_{\mu\nu}\equiv\delta
T_{\mu\nu}\equiv T_{\mu \nu}-\bar{T}_{\mu\nu}$ induced by the fluctuations $
h_{\mu\nu}$. In first order 
\begin{equation}
\partial^{\mu}\tau_{\mu\nu}-\frac{1}{2}T_{\mu\alpha}^{(0)}\,\partial_{\nu
}h^{\mu\alpha}=0\;, \label{sete}
\end{equation}
where $T_{\mu\nu}^{(0)}=\mathrm{diag}(\rho_{0},p_{0},p_{0},p_{0})$\ replaces 
$\bar{T}_{\mu\nu}$, an approximation that will be made whenever $\bar{T}
_{\mu\nu}$\ is multiplied by the field $h_{\alpha\beta}$ because $\bar{T}
_{\mu\nu}-T_{\mu\nu}^{(0)}$ is already suppressed by $\bar{h}_{\alpha\beta}$
. Equations (\ref{sete}) give the linearized equations (where, $\ \dot{}
\equiv\partial/\partial t$),
\begin{align}
\delta \dot{\rho}-(\rho _{0}+p_{0})\,u_{i,i}+\frac{1}{2}(\rho _{0}+p_{0})\,
\dot{h}_{00}& =0\;, \label{oitoa} \\
(\rho _{0}+p_{0})\,\dot{u}_{i}-\delta p_{,i}-\frac{1}{2}(\rho
_{0}+p_{0})\,h_{00,i}& =0\;, \label{oitob}
\end{align}
which yield $\tau _{\mu \nu }$\ as linear functions of $h_{\alpha \beta }$
when considering periodic waves. This system is completed by Einstein
equations (\ref{dois}) and (\ref{quatro}) resulting in the approximate
equality, 
\begin{equation}
R_{\mu \nu }[\bar{h}+h]-R_{\mu \nu }[\bar{h}]=8\pi G\left( \tau _{\mu \nu }-
\tfrac{1}{2}\eta _{\mu \nu }(\tau -T_{\alpha \beta }^{(0)}\,h^{\alpha \beta
})-\tfrac{1}{2}T^{(0)}h_{\mu \nu }\right) \;, \label{nove}
\end{equation}
where the traces are defined as $\tau =\tau _{\mu \nu }\,\eta ^{\mu \nu }$\
and $T^{(0)}=T_{\mu \nu }^{(0)}\,\eta ^{\mu \nu }$. In the right-hand side
we neglect terms that go as, omiting the indices, $T\bar{h}h$, $(\bar{T}
-T^{(0)})h$ or $\tau \bar{h}$, because they are suppressed by extra factors
of $\bar{h}_{\mu \nu }$ and henceforth only appear at $G^{2}$ order.

In the weak field limit the left-hand side of Eq.~(\ref{nove}) is
approximated by $-\frac{1}{2}\square h_{\mu \nu }$ whereas in the right hand
side the explicit dependence of the matter terms in $h_{\mu \nu }$ is
neglected. In the study of the propagation of gravitational waves through
matter one usually considers the high frequency regime, where the wavenumber 
$k$ is much greater than the background curvature, i.e., $k^{2}\gg 8\pi
G\rho $, enabling one to neglect in the equations the matter density $\rho$. 
However, matter effects become important when $k^{2}$ approaches the
background curvature, i.e., $k^{2}\sim 8\pi G\rho $. For such small $k$
higher order terms of the Ricci tensor, such as $\triangle \bar{h}
_{00}\,h_{\mu \nu }\simeq 4\pi G\rho \,h_{\mu \nu }$, are of the same order
as $k^{2}h_{\mu \nu }$ and still linear in $h_{\mu \nu }$. This calls for a
higher order expansion of the Ricci tensor in equation (\ref{nove}) up to
terms linear in both $h_{\mu \nu }$ and background field $\bar{h}_{\mu \nu}$. 
Such an expansion includes products of $\bar{h}_{\mu \nu }$ derivatives
and $h_{\mu \nu }$ derivatives and products of $\bar{h}_{\mu \nu }$ times
second derivatives of $h_{\mu \nu }$ so that translational invariance is
lost and monochromatic waves cease to be in general solutions of Einstein
equations. This may be circumvented in some cases, such as in the solutions
presented below.

Consider a thin wall of fluid lying over the plane $x=0$ with
translational symmetry along the $y-z$ plane. If the pressure is
everywhere much smaller than the density the Einstein equations
(\ref{quatro}) admit in first order a Newtonian solution $\bar{h}_{\mu
\nu }=2\phi (x)\,\delta _{\mu \nu }$ with $\phi ^{\prime \prime }=4\pi
G\bar{\rho}$. Consider in addition that the fluid is in a nearly
incompressible state, i.e., with a practically uniform density
$\bar{\rho}\simeq \rho _{0}$. A solution with reflection symmetry
about the plane $x=0$ possess an acceleration $\phi ^{\prime }=4\pi
G\rho _{0}x$ and pressure $\bar{p}=2\pi G\rho _{0}^{2}(\ell
^{2}/4-x^{2})$, $\bar{p }$ vanishing at the boundary planes $x=\pm
\ell /2$. The density variation is negligible across the wall if
$\Delta \bar{p}\ll c_{s}^{2}\rho _{0}$ for a given speed of sound
$c_{s}$. That is true if the wall thickness $\ell $ obeys $\ell \ll
\lambda _{J}$, where $\lambda _{J}=\sqrt{\pi {c_{s}^{2}} /G\rho }$ is
the Jeans wavelength. Note that the non-relativistic condition $
\bar{p}\ll \bar{\rho}$ is a weaker condition since it implies $\ell
\ll 1/ \sqrt{G\rho }$. The magnitude of the potential $\phi $ at the
wall depends upon the longitudinal dimension of the wall, $L$, because
at distances larger than $L$ it decays with the inverse distance to
the wall. One obtains $|\phi |\sim G\rho _{0}\ell L$ at the
wall. Therefore the Newtonian condition $\phi \ll 1$ yields $\ell L\ll
1/G\rho $. To summarize, the background we are assuming consists of a
fluid with uniform density and small pressure associated with a
Newtonian gravity potential. Other details, such as the precise fall
off of the pressure at the rim of the wall, are no needed here.

To obtain from Eq.~(\ref{nove}) an accurate system of linear equations in $
h_{\mu\nu}$ up to order of $G\rho$ it is sufficient to have $\bar{h}_{\mu\nu}
$ in leading order of $G\rho$ in the left-hand side of Eqs.~(\ref{nove}).
There are terms proportional to $\phi^{\prime\prime }h_{\mu\nu}$, $
\phi^{\prime}h_{\mu\nu,\alpha}$ and $\phi\,h_{\mu\nu ,\alpha\beta}$. For
definiteness we consider waves propagating along the $z$ axis, i.e., $
\tau_{\mu\nu}$ and $h_{\mu\nu}$ varying as $\exp(-i\omega t+ikz)$. It is
conceivable that for large enough wavelengths the terms proportional to $
\phi^{\prime}$ and $\phi$ are much smaller than the terms with $
\phi^{\prime\prime}=4\pi G\rho_{0}$ so that solutions of that type may be
found by neglecting the $\phi^{\prime}$ and $\phi$ terms. It is not enough
however that $k^{2}\phi$ and $k\phi^{\prime}$ be much smaller than $
\phi^{\prime\prime}$. The three kinds of terms mix in the Bianchi identities
and therefore one should not expect that the Einstein equations remain
always consistent with each other if some terms are neglected. Nevertheless
two modes were found that provide consistent solutions of Einstein equations
with definite wavelength along the $z$ axis. They are $h_{12}$ and $
h_{02}-h_{23}$.

After neglecting the $\phi ^{\prime }$ and $\phi $ terms (and pressure)
Eqs.~(\ref{nove}) give for the $h_{12}$ mode 
\begin{equation}
(\omega ^{2}-k^{2}-2\phi ^{\prime \prime })h_{12}=16\pi G(\tau _{12}-\tfrac{1
}{2}\rho _{0}h_{12})\;, \label{treze}
\end{equation}
while the fluid equations of motion (\ref{oitoa}) and (\ref{oitob}) yield $
\tau _{\mu \nu }=0$. Hence, the $h_{12}$ mode satisfies $\omega ^{2}=k^{2}$
and propagates at the vacuum speed of light without disturbing the fluid.
This absence of dispersion is in agreement with various previous results
regarding the propagation of transverse gravitational waves in media (see,
e.g., \cite{polnarev,grishchuk}).

For the $h_{02}$ and $h_{23}$ mode ($kh_{23}=-\omega h_{02}$ in the gauge 
(\ref{cinco})) Eqs.~(\ref{nove}) give, after neglecting the $\phi^{\prime}$
and $\phi$ terms as before: 
\begin{align}
(\omega^{2}-k^{2})h_{02} & =-16\pi G(\delta T^{02}-\tfrac{1}{2}\rho_{0}
h_{02})\;, \label{catorzea} \\
(\omega^{2}-k^{2})h_{23} & =16\pi G(\delta T^{23}-\tfrac{1}{2}\rho_{0}
h_{23})\;, \label{catorzeb}
\end{align}
where $\delta T^{02}=\rho_{0}u^{2}$ is the momentum density along the $y-$
direction and the fluid velocity is $u^{2}=-u_{2}+h_{02}$. The terms
proportional to $\rho_{0}h_{02}$ and $\rho_{0}h_{23}$ are due to the
background curvature $G\rho_{0}$. The fluid equations of motion (\ref{oitoa}) 
and (\ref{oitob}) yield $u_{i}=0=\delta\rho=\delta p$ hence, $u^{2}=h_{02}$
and $\delta T^{23}=0$. The resulting dispersion relation is 
\begin{equation}
\omega^{2}=k^{2}-8\pi G\rho_{0}\;. \label{quinze}
\end{equation}
In contrast to the $h_{12}$, this $h_{02}-h_{23}$ mode exhibits an
instability at very large wavelengths, $k^{2}\leq8\pi G\rho_{0}$, that is
induced by matter. Most notably, the polarization carries the helicity
states $\pm1$ and not $\pm2$, in contrast to the gravitational waves that
propagate in vacuum. In vacuum, of course, this mode can be gauged away
through a coordinate transformation but that is no longer true in matter as
can be seen by the following argument: $h_{23}$ can be eliminated with a
transformation of the type $y\rightarrow y+\epsilon(t,z)$ but $h_{02}$
transforms into $h_{02}^{\prime}=(1-\omega^{2}/k^{2})h_{02}$, which is not
zero, because the phase velocity $\omega/k$ is different from unit in
matter. Alternatively, one may gauge away $u^{2}\ $and $h_{02}$ but not $
h_{23}$. It is easy to prove that no coordinate transformation can bring
simultaneously all the components $h_{\mu\nu}$ and $u^{i}$ into zero.

The motion of the fluid can be characterized in a gauge independent way by
the covariant derivative of the velocity, $u_{\mu ;\nu }$ (see \cite{ellis89}).
In leading order, the important non-zero covariant derivatives are 
\begin{equation}
u_{2;3}=u_{3;2}=-\frac{1}{2}(\partial _{3}h_{02}-\partial _{0}h_{23})\;.
\label{u23}
\end{equation}
It means that this is a shear motion, $u_{\mu ;\nu }=\sigma _{\mu \nu }$,
with the square of the shear magnitude defined as the gauge invariant
quantity $\sigma ^{2}=\sigma _{\mu \nu }\sigma^{\mu \nu }/2$. In the gauge
we use, 
\begin{equation}
\sigma =u_{2;3}=-\frac{4\pi G\rho _{0}}{k^{2}}\,\partial _{3}h_{02}\;,
\label{sigma}
\end{equation}
which shows again the crucial role played by matter. These helicity-1 
waves can then be called shear waves. The components $u^2$, $h_{02}$, 
$h_{23}$ correspond to what are called vector perturbations in 
cosmological perturbation theory \cite{liddlelyth}. They are interpreted 
as a relativistic generalization of purely rotational fluid flow. Here, 
they appear exclusively associated to pure shear motion. The vorticity 
for this mode is zero, since its relevant covariant velocity component 
$u_2=0$ (indeed, from Euler equations: $u_2=-u^2+h_{02}=0$).

We can now discuss the two main aspects of this problem: the first aspect
concerns the physics of the instability, and the second the generation and
propagation of stable waves.

\begin{figure}[t]
\centerline{\epsffile{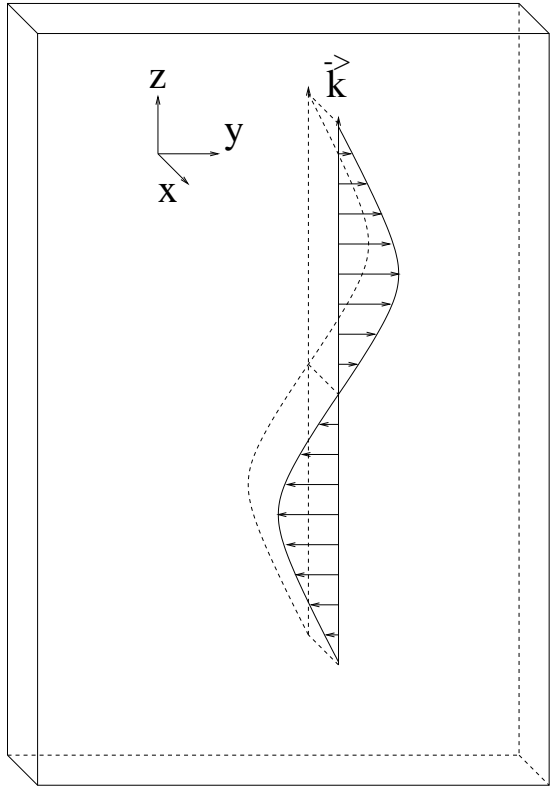}} \vskip0.5cm
\caption{A helicity-1 shear wave of gravity travels in the
$z-$direction through a plane parallel slab and induces a motion on
the fluid in the $y-$direction. The static gravitational potential is
a function of the $x$ coordinate.}
\label{fig1}
\end{figure}

Starting with the first aspect, one notices from equation (\ref{quinze}),
that the instability sets in through an $\omega =0$ mode, with a critical
wave number $k_{c}=\sqrt{8\pi G\rho _{0}/c^{2}}$, where the speed of light $c
$ has been restored. In terms of wavelength, above the critical value $
\lambda _{c}=2\pi /k_{c}$, i.e., for 
\begin{equation}
\lambda \geq \left( \frac{\pi c^{2}}{2G\rho _{0}}\right) ^{1/2}\;,
\label{dezasseis}
\end{equation}
the system is unstable, $\lambda =\lambda _{c}$ being the marginal case.
This instability tends to shear in the material along the node of the
perturbation. Indeed, under the effect of the  wave the particles of the
fluid get an acceleration $\Gamma _{200}=\dot{h}_{02}$ that produces the
velocity $u^{2}=h_{02}$. As shown in Figure~\ref{fig1}, one sees that part
of the material moves to one side, part to the other, whereas it stays at
rest at the node where the velocity in the $y-$direction, $u^{2}$, is zero.

A very important point which should be made explicitly clear is that at such
critical wavelengths there are equilibrium configurations which are not at
all dynamically unstable. Indeed, in a previous paragraph we have put the
following limits on the equilibrium configuration, $\ell \ll c_{s}/\sqrt{
G\rho }$ and $\ell L\ll c^{2}/G\rho $. To have $L>\lambda _{c}\sim c/
\sqrt{G\rho }$ while the above conditions are still satisfied, $\ell $ has
to satisfy both limits $\ell \ll c_{s}\lambda _{c}/c$ and $\ell \ll \lambda
_{c}^{2}/L$. Thus, for sufficiently small thickness $\ell $, there are well
defined equilibria.

%
%
%
%
\ 
In order to better understand such an instability it is advisable to
rearrange equation ({\ref{dezasseis}}) above and write it as $\mu \geq \pi
m_{\ell }/\lambda $, where $\mu \equiv \sigma \lambda \equiv \rho _{0}\ell
\lambda $ is the mass per unit length in the $y-$direction and $m_{_{\ell
}}\equiv c^{2}\ell /2G$ is the reference relativistic mass associated with
the thickness length $\ell $. Thus $m_{\ell }/\lambda $ can be seen as the
relativistic reference mass per unit length in the $z-$direction. The
instability criterion ({\ref{dezasseis}}) says that when the mass per unit
length in the $y-$direction of the perturbation is about greater than the
reference mass per unit length in the $z-$direction then the momentum in the 
$y-$direction is enough to overcome the pull back from the mass along $z$
and the material breaks apart.

This instability is of course different from Jeans instability but not quite
in what regards the gravitational field dynamics. Equation (\ref{catorzea})
can be written as $\square h_{02}=8\pi G\,\delta T^{02}$, because $
u^{2}=h_{02}$, and if one performs a Lorentz boost along the $y-$direction,
it is equivalent to $\square h_{00}^{\prime }=-8\pi G\,\delta T^{\prime 00}\;
$in the new frame, which is comparable to the equation governing the
response of the gravitational field to density fluctuations. What is
different is that they are not accompanied by perturbations in the fluid
proper density and pressure as in the Jeans, acoustic case, but rather by
transverse displacements of the fluid. For such a plane layer and density
wave perturbations of the form $\exp (-i\omega t+ikz)$, Jeans instability
sets in when $k\leq k_{J}$ with $k_{J}=\sqrt{4\pi G\rho _{0}/{c_{s}^{2}}}$,
or $\lambda \geq \lambda _{J}$ with $\lambda _{J}=\sqrt{\pi {c_{s}}
^{2}/G\rho _{0}}$. The material collapses into a set of plane-parallel slabs
distributed perpendicularly to the $z-$direction and uniformly along the $z-$
direction \cite{kippenhahnweigert,spitzer}.

The instability we have found here sets in through the interaction of
shear waves and matter. A helicity-1 shear wave travelling through the
plane parallel layer in the $z-$direction with $\lambda\geq\lambda_{c}$ will
induce an instability in the matter and in the gravitational field itself.
For a planar slab characterized by a surface density $\sigma$ and height $
\ell$ one can write the critical wavelength as $\lambda_{c}=\sqrt{\pi c^{2}
\ell/2G\sigma}$. One sees that $\lambda_{c}$ can be relatively small for
high enough surface densities or very thin systems. Conversely, if such
instabilities are not detectable within the system, one can put a lower
limit on the ratio $\ell/\sigma$; for a slab with size $L$ these
instabilities do not show up if $\lambda_{c} \gtrsim L$, i.e., $
\ell/\sigma\gtrsim(2G/\pi c^{2}) L^{2}$. An estimate of $\lambda_{c}$ for
our Galaxy yields $\lambda_{c}\simeq16\,000 \,$Kpc showing that it is not
thin enough to present such instabilities. However, for extended and very
thin systems this instability is applicable and could set in.

Within plane parallel symmetric configurations, other unstable modes should
occur, although we have been able to solve the system consistently only for
the $h_{02}-h_{23}$ mode because we have restricted this study to plane
waves. It would be interesting to find a consistent solution for the mode
involving $h_{00}$. 

We have considered here plane parallel symmetry but other type of
configurations, such as disk-like, should also possess such
instabilities. For disk-like configurations some other features and
new global instabilities could set in. Jeans instabilities show
surprises in the disk-like case, since for high enough values of the
sound speed a disk is stable to all axisymmetric perturbations
\cite{toomre,goldreichlyndenbell}, a feature that does not appear in
the spherical and planar cases. Domain walls and cosmic strings should
be considered. In a preliminary calculation, we have found that domain
walls also present this type of instability to wave perturbations of
the gravity and scalar field \cite{bl00}.  On the other hand, the
spherical case is less interesting since here $\lambda _{c}\sim
R^{3/2}/R_{s}^{1/2}$, where $R_{s}$ is the Schwarzschild radius of an object
of radius $R$. Thus, the instability sets in when $R\sim R_s \sim
\lambda _{c}$, and other stronger dynamical instabilities must already
be present. 

We now discuss the second aspect, i.e., the generation and propagation of
stable waves of helicity-1. In order to estimate the energy
flux of these waves we use the expression for the effective
stress-energy $t_{\mu \nu }$ tensor given in \cite{mtw} and apply to our
case. It yields 
\begin{equation}
t_{\mu \nu }=\frac{1}{16\pi G}<-h_{02,\mu }h_{02,\nu }+h_{23,\mu }h_{23,\nu
}>=-\frac{\rho _{0}}{2k^{2}}<h_{02,\mu }h_{02,\nu }>\;, \label{flux1}
\end{equation}
where the symbol $<\;>$ denotes an average over one wave period. Defining 
$u^{2}=|u|\exp (-i\omega t+ikz)$, with $|u|$ being the amplitude of the
velocity in the $y-$direction, we find 
\begin{equation}
t^{03}=\frac{1}{2}\rho _{0}|u|^{2}v\;, \label{flux2}
\end{equation}
where $v$ is the velocity of propagation of the wave. It is now difficult to
estimate at this point a realistic energy flux, given one has to deal with
the propagation as well as the generation of such a wave. One possibility is
to have a source of mass $M$ in an oscillatory (orbital) movement of radius $
R$, inducing tidal motions on particles at a distance $r$. This induces a
tidal velocity to the particles which we identify with $|u|$, given then by $
|u|/c\sim \sqrt{R_{s}R}/r$, where $R_{s}=2GM/c^{2}$. 
If we put $R_{s}\sim 10^{-6}R$, $R\sim 10^{-2}r$ and $\rho_0 = 1\,$g/cm$^3$
we obtain $|u|/c\sim 10^{-5}$ and, for $v\sim
c$, an energy flux $t^{03}\sim 10^{20}\,$erg/(s$\,$cm$^{2}$). Perhaps, the
most important point is to notice that this tidal perturbation
propagates away through the matter at typically the speed of light according
to the dispersion relation (\ref{quinze}) and properties we have derived for
this helicity-1 wave. One can then think that randomly
distributed waves of the type considered, should induce
thermal motions in the matter, which would be detected as velocity
dispersions in the stars or particles of the system. These velocities, in
turn, would give the impression of the existence of an extra Newtonian
gravitational potential which, for instance, could be interpreted as coming
from dark matter, or from modifications of the gravitational potential
itself.

The propagation of these helicity-1 shear waves and the instabilities
themselves we have been studying here can be considered wave
detectors in potential and deserve close attention.

\vskip1cm

\section*{Acknowledgments}

This work was partially funded by FCT through project ESO/PRO/1250/98.

\end{document}